\documentclass[letterpaper,11pt,oneside,onecolumn]{article}
\usepackage{amsmath, amssymb, amstext}
\usepackage{amsthm}
\usepackage{csquotes}

\usepackage{mathtools}
\usepackage{authblk}
\usepackage[colorlinks]{hyperref}
\usepackage{enumerate}  
\usepackage{listings}
\usepackage{centernot}

\newcommand{\M}{\mathcal{M}}
\newcommand{\B}{\mathcal{B}}

\DeclareMathOperator{\ordOp}{ord}
\DeclareMathOperator{\stabOp}{stab}
\DeclareMathOperator{\AMO}{AMO}

\newcommand{\Formula}{\mathcal{F}}
\newcommand{\Formulay}{\mathcal{F}'}
\newcommand{\Ford}{\mathcal{F}_{\ordOp}}
\newcommand{\Fz}{\mathcal{F}_z}
\newcommand{\Fstab}{\mathcal{F}_{\stabOp}}
\newcommand{\Fstaby}{\mathcal{F}_{\stabOp}'}
\newcommand{\Model}{\mathcal{F}^+}
\newcommand{\Modely}{{\mathcal{F}^+}'}

\newtheorem{theorem}{Theorem}

\newtheorem{remark}[theorem]{Remark}

\newtheorem{conj}[theorem]{Conjecture}

\author{Kanstantsin Pashkovich}
\author{Laurent Poirrier}

\title{Three-dimensional Stable Matching with Cyclic Preferences}

\affil{Department of Combinatorics and Optimization,\\ University of Waterloo,\\ 200 University Avenue West,
Waterloo, ON, Canada N2L 3G1\\
  \texttt{\{kpashkov,lpoirrier\}@uwaterloo.ca}}

\begin{document}

\maketitle 

\begin{abstract}
We consider the three-dimensional stable matching problem with cyclic preferences, a problem originally proposed by Knuth. Despite extensive study of the problem by experts from different areas, the question of whether every instance of this problem admits a stable matching remains unanswered. In 2004, Boros, Gurvich, Jaslar and Krasner showed that a stable matching always exists when the number of agents in each of the groups is three. In 2006, Eriksson, Sj\"{o}strand and Strimling showed that a stable matching exists also when the number of agents in each group is four. In this paper, we demonstrate that a stable matching exists when each group has five agents. Furthermore, we show that there are at least two distinct stable matchings in that setting.
\end{abstract}

\section{Introduction}
The stable matching problem is one of the most important problems in algorithmic game theory. In the stable matching problem, we are given two disjoint sets of agents, where each agent in one set has a preference order over agents in the other set. We are interested in finding a matching of agents, i.e. we are interested in pairing some agents from the first set with some agent from another set. Moreover, we want the resulting matching to be stable, i.e. there should exist no two agents which prefer each other to their assigned partners. In their seminal paper, Gale and Shapley~\cite{Gale} show that a stable matching always exists and they provide an algorithm for finding a stable matching in polynomial running time.

The situations changes when we have three disjoint groups of agents, where the agents in the first, second and third groups have a preference order over the agents in the second, third and first groups, respectively.  In an instance of the \emph{three-dimensional stable matching problem with cyclic preferences} (c3DSM), we have three disjoint sets of agents $A$, $B$ and $C$ such that $|A|=|B|=|C|=n$. Each agent $a$ in the set $A$ has a total order of the agents in $B$ according to the preference of $a$. Thus, given $a\in A$ and $b,b'\in B$ we can compare the agents $b$ and $b'$ according to the preference of $a$. For the sake of exposition, we use the notation $b >_a b'$ to say that $a$ prefers $b$ to $b'$. Similarly, each agent $b$ in the set $B$ has a total order of the agents in $C$; and each agent $c$ in the set $C$ has a total order of the agents in~$A$.

A (perfect) \emph{matching} $M$ is a set of $n$ triples of the form $(a,b,c)$, $a\in A$, $b\in B$ and $c\in C$. Given a matching $M$, for all $a\in A$ we denote by $M(a)$ the agent $b\in B$ such that $M$ contains a triple of the form $(a,b,c)$ for some $c\in C$. Similarly, for $b\in B$ we denote by $M(b)$ the agent $c\in C$ such that $M$ contains a triple of the form $(a,b,c)$ for some $a\in A$; for $c\in C$ we denote by $M(c)$ the agent $a\in A$ such that $M$ contains a triple of the form $(a,b,c)$ for some $b\in B$.

 A triple $(a,b,c)$, $a\in A$, $b\in B$, $c\in C$ is called \emph{blocking} for a matching~$M$ if
\[
b>_a M(a)\quad\text{and}\quad c>_b M(b)\quad\text{and}\quad a>_c M(c)\,.
\]
A matching $M$ is called \emph{stable} if there exists no blocking triple for $M$. The following conjecture generalizes the results of Gale and Shapley~\cite{Gale} and is usually attributed to Knuth.
\begin{conj}
Every instance of (c3DSM) admits a stable matching.
\label{conj:knuth}
\end{conj}

Eriksson, Sj\"{o}strand and Strimling further strengthened Conjecture~\ref{conj:knuth} as follows.
\begin{conj}
Every instance of (c3DSM) admits at least two distinct stable matchings.
\label{conj:eriksson}
\end{conj}

\subsection{Previous Work}

The study of the generalization of the stable matching problem to three groups was initiated by Knuth~\cite{Knuth} (Problem 11, page 97):
\begin{displayquote}
\emph{Peut-on g\'en\'eraliser le probl\`eme du couplage stable \`a trois ensembles d'objects (par example hommes, femmes, chiens)?}
\end{displayquote}
The introduction of \emph{cyclic} (or ``circular'') preferences (c3DSM), was attributed to Knuth by one anonymous referee in the paper of Ng and Hirschberg~\cite{Ng} on three-dimensional stable matching problems:
\begin{displayquote}
\emph{One of the referees, who called our attention to this problem, attributes its origin to Knuth and ``dubbed" it circular 3GSM. The complexity of this problem is currently an open question.}
\end{displayquote}
 In 2004, Boros, Gurvich, Jaslar and Krasner showed that every instance of (c3DSM) with $n=3$ admits a stable matching~\cite{Boros}. In 2006, Eriksson, Sj\"{o}strand and Strimling showed that a stable matching always exists also when $n=4$~\cite{Eriksson}. The approach of Eriksson, Sj\"{o}strand and Strimling is based on a strengthening of the result for $n= 3$ and then exploiting this strengthening to show that Conjecture~\ref{conj:knuth} (and~\ref{conj:eriksson}) can be answered positively for~$n\leq 4$.

Bir\'o and McDermid showed that for the variant of the problem in which some agents are \emph{not acceptable} for some agents of the previous group, a stable matching might not exist~\cite{Biro}. Moreover, in this case the decision problem about the existence of a stable matching becomes $\mathcal{NP}$-complete~\cite{Biro}. Farczadi, Georgiou, and K\"onemann  showed that it is $\mathcal{NP}$-complete to decide whether a given partial matching consisting of pairs $(a,b)$, $a\in A$, $b\in B$ can be extended to a stable matching consisting of triples $(a,b,c)$, $a\in A$, $b\in B$ and $c\in C$~\cite{Farczadi}.

\subsection{Our Results}
In this paper, we follow the natural approach and formulate the verification of Conjecture~\ref{conj:knuth} (then~\ref{conj:eriksson}) as a satisfiability problem, obtaining the following theorems.

\begin{theorem}
For $n=5$ every instance of (c3DSM) admits a stable matching.
\end{theorem}
\begin{theorem}
For $n=5$ every instance of (c3DSM) admits at least two distinct stable matchings.
\end{theorem}

We would like to remark that a naive inspection approach does not turn out to be fruitful for this problem. This is due to the large number of instances of (c3DSM). For example, for $n=5$, the number of (labeled) instances of (c3DSM) is $(5!)^{15}\approx 1.54\cdot 10^{31}$. After the elimination of symmetries, we still have at least $(5!)^{12}/ 3\approx 2.97\cdot  10^{24}$ possible instances. For $n=6$ these numbers look as follows: $(6!)^{18}\approx 2.7\cdot  10^{51}$ and $(6!)^{15}/3\approx 2.42\cdot  10^{42}$, explaining why it is hard to verify Conjecture~\ref{conj:knuth} using a naive inspection approach.

\section{Our Approach}
\label{sect:formula}

To prove that Conjecture~\ref{conj:knuth} can be answered positively for $n=5$, we formulate a satisfiability problem such that there exists an instance of (c3DSM) with no stable matching if and only if the constructed formula is satisfiable.

For $a\in A$, $b,b'\in B$, $b\neq b'$, we let the variable $x_{a,b,b'}$ indicate the preference of $a$ between $b$ and $b'$. Specifically, $x_{a,b,b'}$ is assigned True if $a$ prefers $b$ to $b'$, i.e., $b >_{a} b'$, and False otherwise. It follows that $x_{a,b,b'} = \lnot x_{a,b',b}$. In our formulation of the satisfiability problem, we use only one of the variables $\{ x_{a,b,b'}, x_{a,b',b} \}$ as an explicit variable while the other is implicitly defined as its negation. However, as a notation, we still use both $x_{a,b,b'}$ and $x_{a,b',b}$ in the following exposition, for the sake of simplicity. We define similarly the variables $x_{b,c,c'}$ for $b\in B$, $c,c'\in C$, $c\neq c'$ and the variables $x_{c,a,a'}$ for $c\in C$, $a,a'\in A$, $a\neq a'$.

Next, we construct a formula~\eqref{eq:formula} such that there exists an instance of (c3DSM) with no stable matching if and only if there exists a value for $x$ satisfying~\eqref{eq:formula}.  Our formula~\eqref{eq:formula} is a conjunction of three formulas~\eqref{eq:formula_order},~\eqref{eq:formula_z} and~\eqref{eq:formula_stab}. The first formula~\eqref{eq:formula_order} captures the requirement that each agent in $A$ (resp. $B$, $C$) has a total order of agents in $B$ (resp. $C$, $A$). This is modeled by the following statements
\begin{alignat*}{9}
&x_{a,b,b'} &&\land x_{a,b',b''} && \implies x_{a,b,b''},\quad&&\forall\, a\in A,\;\;&b,&b',&b''&\in B,\;\;&& b&\neq b'&\neq b''&&\neq b,\\
&x_{b,c,c'} &&\land x_{b,c',c''} && \implies x_{b,c,c''},\quad&&\forall\, b\in B,\;\;&c,&c',&c''&\in C,\;\;&& c&\neq c'&\neq c''&&\neq c,\\
&x_{c,a,a'} &&\land x_{c,a',a''} && \implies x_{c,a,a''},\quad&&\forall\, c\in C,\;\;&a,&a',&a''&\in A,\;\;&& a&\neq a'&\neq a''&&\neq a,
\end{alignat*}
which can be written in conjunctive normal form (CNF) as
\begin{alignat*}{2}
&\bigwedge_{\substack{a\in A\\b, b', b''\in B\\b\neq b'\neq b''\neq b}}&&(\lnot x_{a,b,b'}\lor \lnot x_{a,b',b''} \lor x_{a,b,b''})\\
\tag{$\Ford$}\label{eq:formula_order}
\land&\bigwedge_{\substack{b\in B\\c, c', c''\in C\\c\neq c'\neq c''\neq c}}&&(\lnot x_{b,c,c'}\lor \lnot x_{b,c',c''} \lor x_{b,c,c''})\\
\land&\bigwedge_{\substack{c\in C\\a, a', a''\in A\\a\neq a'\neq a''\neq a}}&&(\lnot x_{c,a,a'}\lor \lnot x_{c,a',a''} \lor x_{c,a,a''}).
\end{alignat*}
Observe that a value of $x$ satisfies~\eqref{eq:formula_order} if and only if it corresponds to an instance of~(c3DSM).

Next, let us denote by $\M$ the set of all possible matchings, and by $\B(M)$ the set of potential blocking triples for a given matching $M \in \M$, i.e., $\B(M) := \{ (a, b, c) \in A \times B \times C \; : \; b\neq M(a), c\neq M(b), a\neq M(c) \}$. In the second formula~\eqref{eq:formula_z}, we introduce auxiliary variables $z_{M,a,b,c}$, for all $M\in \M$ and $(a,b,c) \in \B(M)$. Each variable $z_{M,a,b,c}$ can be True only if the triple $(a,b,c)$ is blocking for the matching $M$, i.e.,
\[
z_{M,a,b,c} \implies x_{a,b,M(a)}\land x_{b,c,M(b)}\land x_{c,a,M(a)},
\]
which is equivalent to
\[\tag{$\Fz$}\label{eq:formula_z}
\bigwedge_{\substack{M \in \M \\ (a,b,c) \in \B(M)}}
(\lnot z_{M,a,b,c} \lor x_{a,b,M(a)})
\land
(\lnot z_{M,a,b,c} \lor x_{b,c,M(b)})
\land
(\lnot z_{M,a,b,c} \lor x_{c,a,M(a)}).
\]
Finally, we define the third formula~\eqref{eq:formula_stab} as follows:
\[\tag{$\Fstab$}\label{eq:formula_stab}
\bigwedge_{M\in \M} \quad \bigvee_{(a,b,c) \in \B(M)} z_{M,a,b,c}\,.
\]
The formula~\eqref{eq:formula_stab} models the condition that, in the instance of (c3DSM) defined by the $x$ variables there exists a blocking triple for every matching in $\M$. In other words, no matching is stable for the instance of (c3DSM) defined by the $x$ variables.

Our resulting formula is defined as follows:
\[\label{eq:formula}\tag{$\Formula$}
\Formula:=\Ford \land \Fz \land \Fstab.
\]

Note that the auxiliary $z$ variables could be eliminated without affecting the satisfiability of~\eqref{eq:formula}. Indeed, the existence of a solution for~\eqref{eq:formula} 
implies the existence of a solution with $z_{M,a,b,c} \iff x_{a,b,M(a)} \land x_{b,c,M(b)} \land x_{c,a,M(a)}$. One could thus set $z_{M,a,b,c} = x_{a,b,M(a)} \land x_{b,c,M(b)} \land x_{c,a,M(a)}$ and substitute out the $z$ variables. However, the introduction of $z$ and~\eqref{eq:formula_z} is one way of putting~\eqref{eq:formula} into CNF.

For Conjecture~\ref{conj:eriksson}, we replace~\eqref{eq:formula_stab} with the alternative formula~\eqref{eq:formula_staby}, in which we associate one additional variable $y_M$ to every clause of~\eqref{eq:formula_stab}.
\[\tag{$\Fstaby$}\label{eq:formula_staby}
\bigwedge_{M\in \M} \quad 
\left( y_M \lor \bigvee_{(a,b,c) \in \B(M)} z_{M,a,b,c} \right).
\]
This models the fact that, for each matching $M$, if $M$ is stable, then $y_M$ is True. Next, we need to express that at most one of the $y_M$ variables is True. \emph{At-most-one} (AMO) constraints are a common feature in SAT formulations, with multiple possible CNF encodings~\cite{Nguyen}. We opted for the so-called \emph{sequential} encoding of Sinz~\cite{Sinz} for its simplicity, and we denote it $\AMO(y_\M)$. We obtain
\[\label{eq:formulay}\tag{$\Formulay$}
\Formulay:=\Ford \land \Fz \land \Fstaby \land \AMO(y_\M).
\]
Observe that~\eqref{eq:formula} is equivalent to fixing every $y_M$ to false in~\eqref{eq:formulay}, in which case $\AMO(y_\M)$ becomes trivially satisfied.

\section{Symmetry Reduction}
\label{sect:sym}
Clearly, the collection of all instances of (c3DSM) possesses many symmetries. If we have an instance of (c3DSM) with no stable matching, then we can construct another instance of (c3DSM) with no stable matching by permuting agents inside the sets $A$, $B$ and $C$ and cyclically permuting the sets $A$, $B$ and $C$. These symmetries can be used to fix the value of some of the $x$ variables, which is equivalent (after SAT preprocessing) to reducing the size of~\eqref{eq:formula}. Before we exploit these symmetries, let us make the following remark which already appeared in~\cite{Eriksson}.
\begin{remark}\label{rem:111-triple}
Assume that, for $n\leq k$, every instance of (c3DSM) admits a stable matching. Then, every instance of (c3DSM) has a stable matching whenever $n=k+1$ and there is a triple $(a,b,c)$, $a\in A$, $b\in B$, $c\in C$ such that
\begin{enumerate}[(i)]
\item $b$ is the most preferred agent for $a$
\item $c$ is the most preferred agent for $b$
\item $a$ is the most preferred agent for $c$.
\end{enumerate}
\end{remark}
\begin{proof}
Let us assume that there is an instance with $n=k+1$  and  with a triple of agents $(a,b,c)$ satisfying the above condition. By the hypothesis, the instance of (c3DSM)  induced on $A\setminus a$, $B\setminus b$, $C\setminus c$ admits a stable matching~$M$. It is straightforward to verify that $M\cup \{(a,b,c)\}$ is a stable matching for the original instance of (c3DSM).
\end{proof}


For the purpose of writing~\eqref{eq:formula}, we let $A=\{a_1,\ldots,a_n\}$, $B=\{b_1,\ldots,b_n\}$ and $C=\{c_1,\ldots,c_n\}$, implicitly assigning an order to the agents. Since (c3DSM) imposes no particular order on the agents within $A$, $B$ or $C$, we can choose one arbitrarily. Therefore, we can fix the order of $B$ according to the preferences of $a_1$ as follows:
\[
b_1 >_{a_1} b_2 >_{a_1} \cdots >_{a_1} b_n,
\]
and fix the order of $C$ according to the preferences of $b_1$:
\[
c_1 >_{b_1} c_2 >_{b_1} \cdots >_{b_1} c_n.
\]
Then, $a_1$ has been determined above, but we can fix the order for $A\setminus \{ a_1 \}$ according to the preferences of $c_1$:
\[
a_2 >_{c_1} \cdots >_{c_1} a_n.
\]
Finally, by Remark~\ref{rem:111-triple} and the result of Eriksson et al.~\cite{Eriksson} for $n=4$, we know that $a_1$ is not the most preferred agent for $c_1$, whenever $n=5$ and the instance of (c3DSM) has no stable matching. For $n=5$, we can thus assume that
\[
a_2 >_{c_1} a_1.
\]

\section{Computation}

The complete SAT model, including~\eqref{eq:formula} from Section~\ref{sect:formula} and the symmetry-breaking constraints from Section~\ref{sect:sym} for $n=5$, is given by the conjunction of the following clauses:
\begin{equation}
\begin{array}{ll}
x_{a_1,b_i,b_j}, & \forall\, 1 \leq i < j \leq n,\\
x_{b_1,c_i,c_j}, & \forall\, 1 \leq i < j \leq n,\\
x_{c_1,a_2,a_1}, & \\
x_{c_1,a_i,a_j}, & \forall\, 2 \leq i < j \leq n,\\
\lnot x_{a,b,b'} \lor \lnot x_{a,b',b''} \lor x_{a,b,b''}, & \forall\, a\in A,\;\; b,b',b'' \in B,\;\; b \neq b' \neq b'' \neq b,\\
\lnot x_{b,c,c'} \lor \lnot x_{b,c',c''} \lor x_{b,c,c''}, & \forall\, b\in B,\;\; c,c',c'' \in C,\;\; c \neq c' \neq c'' \neq c,\\
\lnot x_{c,a,a'} \lor \lnot x_{c,a',a''} \lor x_{c,a,a''}, & \forall\, c\in C,\;\; a,a',a'' \in A,\;\; a \neq a' \neq a'' \neq a,\\
\lnot z_{M,a,b,c} \lor x_{a,b,M(a)}, & \forall\, M \in \M,\;\; (a,b,c) \in \B(M),\\
\lnot z_{M,a,b,c} \lor x_{b,c,M(b)}, & \forall\, M \in \M,\;\; (a,b,c) \in \B(M),\\
\lnot z_{M,a,b,c} \lor x_{c,a,M(a)}, & \forall\, M \in \M,\;\; (a,b,c) \in \B(M),\\
\bigvee_{(a,b,c) \in \B(M)} z_{M,a,b,c}, & \forall\, M \in \M.\\
\end{array}
\tag{$\Model$}
\label{eq:model}
\end{equation}

The \textit{de facto} standard for the input of high-performance SAT solvers is the so-called DIMACS format. It describes a ``flat'' CNF formula, i.e., one where the $v$ variables are numbered $1$ to $v$, and where all the clauses are given explicitly. In order to solve a model such as~\eqref{eq:model}, it is customary to first write it in a domain-specific language (DSL) that allows for the use of sets, quantifiers, and multi-dimensional variable indices. Then, a specialized tool translates it into the desired flat format. Instead, we wrote a small C++ program~\cite{Pashkovich1} whose output is the desired CNF formula in the DIMACS format. We chose this approach for two reasons. First, to the best of our knowledge, all suitable modeling DSLs target constraint programming solvers. However, since~\eqref{eq:model} is a SAT problem, we can use SAT solvers, which are less general but faster. Secondly, we use a small custom utility library that lets us keep the model short and transparently handles the anti-symmetric indexing of $x$ in which $x_{i,j,j'} = \lnot x_{i,j',j}$.

The resulting CNF formula has 864,150 variables and 2,607,327 clauses, and it is unsatisfiable. We solved it with \texttt{plingeling}~\cite{Biere2} on large computer (4x Intel Xeon E5-4660v3, total 112 threads and 256GB of RAM) in about two and a half days (129 CPU days). We confirmed our result with \texttt{glucose-syrup 4.0}~\cite{Een,Audemard2} on a desktop computer (Intel Core i3-6100 with 4 threads and 8GB of RAM) in a bit under 5 days (18 CPU days). Detailed solver output is provided in~\cite{Pashkovich2}.

For the stronger Conjecture~\ref{conj:eriksson}, we have a formulation $\Modely$ that incorporates the symmetry-breaking constraints from Section~\ref{sect:sym} and the AMO encoding from~\eqref{eq:formulay}. Its CNF has 892,948 variables and 2,650,522 clauses, and it is unsatisfiable as well. We solved it with \texttt{plingeling} on the large computer mentioned above in about 9 days (515 CPU days), and confirmed the result with \texttt{glucose-syrup} on the desktop computer in about 32 days (64 CPU days).

\section{Discussion}

We took the straightforward approach of formulating the verification problem for Conjectures~\ref{conj:knuth} and~\ref{conj:eriksson} as SAT problems. While simple, this method lets us answer the questions for $n=5$, which had been open for 12 years. We note that the same method can also provide an alternative proof for $n<5$: \texttt{glucose} takes about 0.12 seconds to prove unsatisfiability of~\eqref{eq:model} for $n=4$, and thus to verify Conjecture~\ref{conj:knuth} for $n=4$. Unfortunately, at the moment we are far from being able to tackle $n=6$, because for $n=6$ our formulation~\eqref{eq:model} has 62,208,270 variables and 187,144,601 clauses.

We thus believe that further theoretical insights are needed in order to reduce the size of~\eqref{eq:model}  and then solve it for $n > 5$. For example, one could try  to follow the approach of~\cite{Eriksson} (see Lemma 2 in~\cite{Eriksson}) and strengthen Conjecture~\ref{conj:knuth} for both $n=4$ and $n=5$. Namely, one can  try to show that for $n=4$ and $n=5$ every instance of (c3DSM) has a stable matching with additional properties, and later use these properties to verify Conjecture~\ref{conj:knuth} for $n>5$ efficiently. At the moment,  we do not have a strengthening of Conjecture~\ref{conj:knuth} for $n>3$ or even a candidate for such a strengthening, which could be used to implement this approach.


\section{Acknowledgements}
We would like to thank Jochen Koenemann and Justin Toth for helpful discussions about the structure of (c3DSM).

\bibliography{bibliography}
\bibliographystyle{plain}

\end{document}